\documentclass{osa-article}
\journal{boe}

\articletype{Research Article}

\usepackage{lineno}

\begin{document}

\pagenumbering{arabic} 
\pagestyle{plain}

\title{Polarization effects on fluorescence emission of zebrafish neurons using light-sheet microscopy}

\author{Hong Ye,\authormark{1} Xin Xu,\authormark{1,2} Jixiang Wang,\authormark{1,2} Jing Wang,\authormark{1,2} Yi He,\authormark{1} Yu Mu,\authormark{3} and Guohua Shi\authormark{1,2,*}}

\address{\authormark{1}Jiangsu Key Laboratory of Medical Optics, Suzhou Institute of Biomedical Engineering and Technology, Chinese Academy of Sciences, Suzhou, China\\
\authormark{2}School of Biomedical Engineering (Suzhou), Division of Life Sciences and Medicine, University of Science and Technology of China, Hefei, China\\
\authormark{3}Institute of Neuroscience, State Key Laboratory of Neuroscience, Center for Excellence in Brain Science and Intelligence Technology, Chinese Academy of Sciences, Shanghai, China}

\email{\authormark{*}ghshi\_lab@sibet.ac.cn} 


\begin{abstract}
Light-sheet fluorescence microscopy (LSFM) makes use of a thin plane of light to optically section and image transparent tissues or organisms {\it{in vivo}}, which has the advantages of fast imaging speed and low phototoxicity. In this paper, we have employed light-sheet microscopy to investigate the polarization effects on fluorescence emission of zebrafish neurons via modifying the electric oscillation orientation of the excitation light. The intensity of the fluorescence emission from the excited zebrafish larvae follows a cosine square function with respect to the polarization state of the excitation light and reveals a 40$\%$ higher fluorescence emission when the polarization orientation is orthogonal to the illumination and detection axes. Through registration and subtraction of fluorescence images under different polarization states, we have demonstrated that most of the enhanced fluorescence signals are from the nerve cells rather than the extracellular substance. This provides us a way to distinguish the cell boundaries and observe the organism structures with improved contrast and resolution.
\end{abstract}

\section{Introduction}

As an advanced imaging approach, light-sheet fluorescence microscopy (LSFM) has been extensively used in developmental biology~\cite{McDole:Cell:859-876, Reichmann:Science:189-193}, cell science~\cite{Liu:Science:284}, and neuroscience~\cite{Hillman:ARN:295-313} in the past decade. Unlike epi- or trans-illumination fluorescence microscopy, LSFM illuminates the sample via a thin laminar laser beam, referred to as a light sheet, in the plane of focus and detects the emitted fluorescence signal in the orthogonal direction~\cite{Huisken:Science:1007-1009, Huisken:Development:1963-75}. This natural feature of optical sectioning brings many salient advantages to the LSFM, such as fast data acquisition, high signal-to-noise ratio, and low phototoxicity and photobleaching, making LSFM the very fit technique for imaging clear biological specimens {\it{in vivo}}~\cite{Santi:JHC:129-138, Olarte:AOP:111-179}.  The zebrafish ({\it{Danio rerio}}), as a vertebrate model organism, shares a high similarity of approximately 70$\%$ with human disease genes~\cite{Santoriello:JCI:2337-2343}, thus playing an important role in many research areas such as pharmacogenetics and neuropharmacology~\cite{Langheinrich:Bioessays:904-12, Kalueff:TIPS:63-75}. The larval zebrafish, due to its transparent property, has become an ideal and popular organism to be studied through light-sheet fluorescence microscopy~\cite{Keller:Science:1065}. 

During a fluorescence imaging process, a fluorescent probe is usually introduced to the living tissue or cells to realize the visualization of fluorescent entities. Due to the photoselection effect induced by the linearly-polarized light, the fluorescence emission is spatially anisotropic~\cite{Hohlbein:JCP:094703, Anantharam:JCB:415-428}. This is because the linear oscillating electric field modulates the randomly distributed molecules into a non-random distribution regarding the orientations of transition dipole moments\cite{Vinegoni:NP:1472-1497}. The molecule with a transition dipole moment oriented in the same direction as the polarized light is excited preferentially and fluoresces primarily on an axis perpendicular to the transition dipole moment, leading to anisotropic fluorescence emission~\cite{Hohlbein:JCP:094703, Lakowicz:2006:Springer}. Since the excitation probability strongly depends on the relative orientation between light polarization and emission dipoles, it allows one to experimentally maximize the fluorescence signal by selecting a specific polarization of excitation light in a microscope. However, if the molecules are light and can rotate without constraint in a medium of low viscosity, the photoselection effect is averaged out. The fluorescence emission will present a certain degree of anisotropy only if the rotation occurs on a time scale longer than the fluorescence lifetime. This happens when the fluorophore is attached to a heavier molecule such as proteins. Therefore, the anisotropic fluorescence emission may reveal the micro-architecture of the biological samples. In case there is a pre-existed anisotropic orientation of the dipole moments that the fluorophores are associated with, the polarization-dependent measurements when using linearly-polarized light can in principle be used to extract information about the spatially-ordered biological structures~\cite{Mojzisova:BPJ:2348-2357}. Therefore, the effects of excitation light polarization on fluorescence emission have attracted enormous interest and been well studied with scanning confocal microscopy~\cite{Steinbach:ACTHIS:316-325, Kress:BPJ:127-136, Micu:Neurophotonics:025002}.

The anisotropic fluorescence emission induced by the polarized excitation light is expected to be more pronounced in the LSFM geometry, where the orthogonal configuration of illumination and detection paths introduces an extra geometrical limitation on the fluorescence signal possible to be collected by the detector, than in the collinear geometry. But few works relevant to the anisotropic fluorescence emission have been carried out in LSFM so far. The polarization-dependent experiment using LSFM has first been conducted to study the contrast mechanism of second harmonic generating (SHG) nanoprobes for {\it{in vivo}} light-sheet imaging~\cite{Malkinson:ACSP:1036-1049}. Recently, the importance of excitation polarization control has been reported via quantitatively demonstrating its influence on the detected fluorescence signal levels in a two-photon LSFM~\cite{Vito:BOE:4651-4665}. However, both works mentioned above involve nonlinear processes, {\it{i.e.}} SHG and two-photon excitation, where the polarization density responds non-linearly to the electric field of the light. This nonlinear effect would complicate the interpretation of images presenting the biophysical phenomena. Detecting anisotropic fluorescence emission using one-photon LSFM can directly map the relationship between light polarization and imaging properties but has not been done experimentally, thus deserves more research.

In this work, we have employed the self-built high-resolution LSFM to study the polarization effects on fluorescence emission using larval zebrafish as the biological specimen. The LSFM system has a lateral resolution of 0.47~$\mu$m, which can resolve the neuronal cells of the zebrafish larvae. We have performed polarization-dependent experiments on both fluorescent microspheres and GCaMP6f-expressed zebrafish larvae. When varying the polarization of the excitation light, the fluorescence intensities obtained from fluorescent microspheres are nearly constant, hardly depending on the polarization states. While in the case of zebrafish larvae, around a 40\% difference in the fluorescence intensity has been observed under different polarization states. With the excitation light propagating along the illumination axis, the fluorescence intensity has a peak value or a valley value when the polarization is perpendicular to or parallel to the detection axis, respectively. To understand the cause of the difference in fluorescence intensity, we have registered the fluorescence images of zebrafish larvae taken under different polarizations on the accuracy of cellular resolution and subtracted them from a reference image. We found that most differential fluorescence signals occurred inside the nerve cells rather than in the extracellular space. The subtraction of fluorescence images of different polarizations eliminates the depolarization fluorescence signal and enables one to obtain a fluorescence image with increased contrast and resolution in a one-photon light-sheet microscope. 

\section{Methods}

\subsection{Sample preparation}

Two types of samples, {\it{i.e.}} fluorescent microspheres and zebrafish larvae, were prepared to study the polarization effects on fluorescence emission in a light-sheet imaging system. The 0.5~$\mu$m fluorescent microspheres used are carboxylate-modified microspheres (F8813, 505/515~nm, Invitrogen). The fluorescent dye molecules are contained inside each polystyrene microsphere. During the experiment, the microspheres were embedded in low-melting-point agarose at a proportion of 1:1000.

Zebrafish larvae of Tg(elavl3:H2B-GCaMP6f) line at 7 days post-fertilization (dpf) were subjected to {\it{in vivo}} imaging. The zebrafish larvae were first immersed in a 10~$\mu$L drop of concentrate bungarotoxin and later filled with 20~mL water for 20 minutes to cut off the synaptic transmission of the neuromuscular junctions. We used the low-melting-point agarose diluted in phosphate-buffered saline (PBS) (1.0$\%$ w/v) for sample fixation. The zebrafish larva was placed in the sample chamber filled with agarose solution at 37~$^{\circ}$C. The position of the zebrafish larva was carefully adjusted before the polymerization of the agarose solution to ensure effective imaging. All the experimental protocols were performed according to the animal experiment guidelines of the Animal Experimentation Ethics Committee of Suzhou Institute of Biomedical Engineering and Technology, Chinese Academy of Science.


\subsection{Experimental setup}

\begin{figure}[htbp]
\centering
\includegraphics[width=\columnwidth]{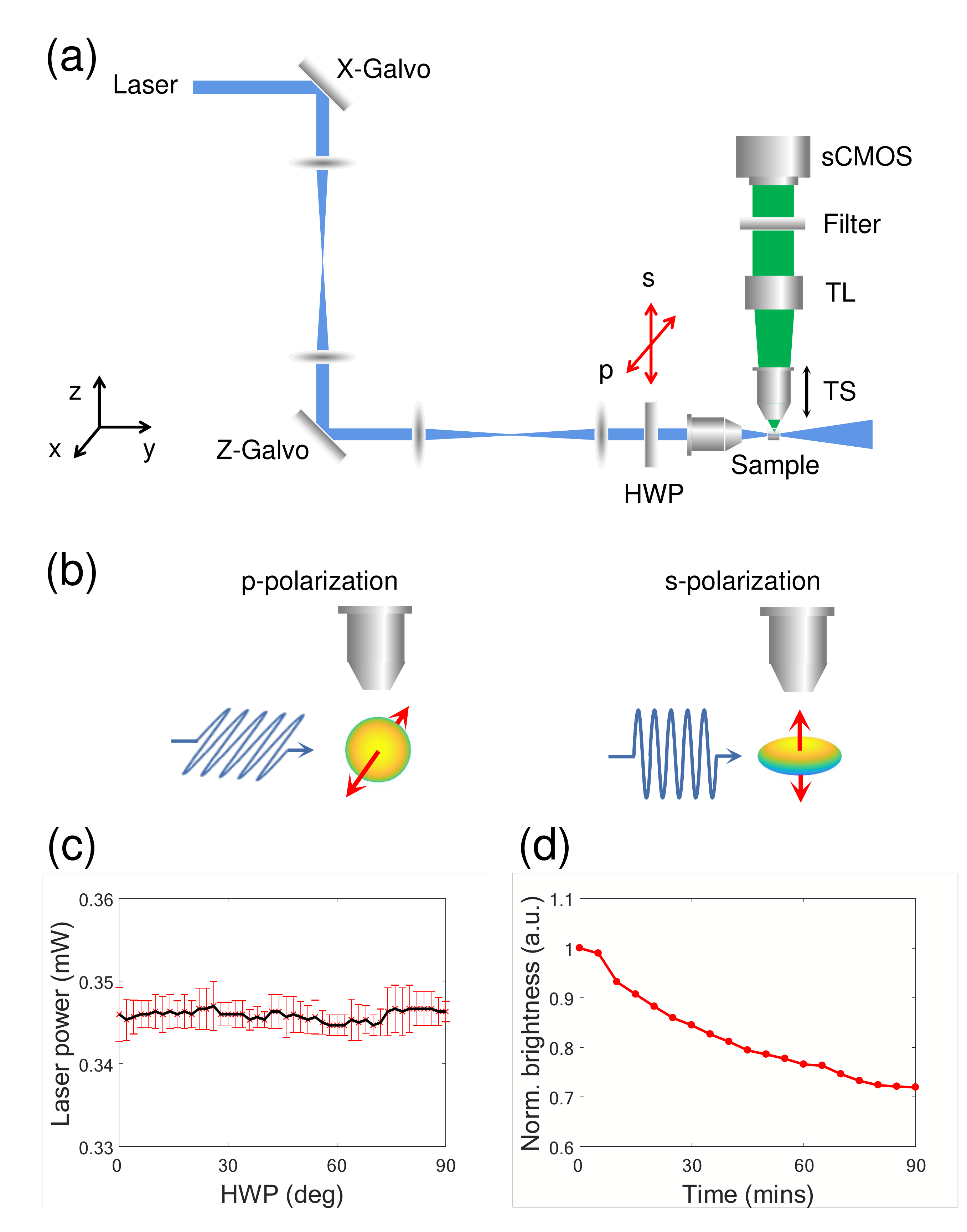}
\caption{(a) Sketch of the LSFM setup with orthogonal illumination and detection axes. HWP: half-wave plate; TL: tube lens; TS: translation stage; (b) The geometry of fluorescence emission with respect to the detection objective when excited with p-polarization (left) and s-polarized (right) light; (c) Power stability of the excitation laser with respect to the polarization; (d) Photobleaching effect as a function of time.}
\label{fig:setup}
\end{figure}

The polarization-dependent experiments were carried out with a self-built one-photon LS microscope. The schematic diagram of the setup is shown in Fig.~\ref{fig:setup}(a), where the illumination (blue color) and detection (green color) paths are orthogonal. A 488~nm wavelength CW laser (W488-25FCB-204, Pavilion Integration Corporation) is employed as the excitation light source. The galvanometer mirrors X-Galvo and Z-Galvo are in conjugate positions with the pupil of the illumination objective (NA=0.14, 5$\times$ Mitutoyo Plan Apochromat Objective). The 2~mm diameter laser beam is scanned by the X-Galvo to form the light sheet in the xy-plane along the larval rostro-caudal direction. Instead of performing the three-dimensional imaging, in our experiments, the Z-Galvo was used to select the lateral xy-plane along the larval dorso-ventral direction to avoid burning the eyes of the zebrafish larvae. The scanning angles of X-Galvo and Z-Galvo largely determine the field of view (FOV) and the imaging depth of the microscopy system, respectively. A high numerical aperture water-immersion detection objective with a magnification factor of 16 (NA=0.8, Nikon CFI75 LWD) is used to collect the emitted fluorescence and followed by a tube lens (TTL200, Thorlabs) to prevent the image quality from degrading, giving a nearly diffraction-limited performance across full FOV of the detection objective while simulated in the commercial software ZEMAX. The position of the detection objective is adjustable by a motorized translation stage so that the focus on the same transverse plane is synchronized with the excitation light sheet when it moves in the z-direction. An sCMOS camera (ORCA-Flash4.0 V3, Hamamatsu) is placed at the end of the detection path to acquire the real-time fluorescence image at a high-speed readout of 100 fps with 2048$\times$2048~pixels. An optical bandpass filter of bandwidth 500-540~nm is placed in front of the sCMOS camera to filter out the green fluorescence and block the stray light. Measurements on fluorescent microspheres of sizes 2~$\mu$m and 0.2~$\mu$m have been carried out to characterize the self-build LSFM system. The microscopy system has a lateral resolution of 0.47~$\mu$m, an axial resolution of 2.98~$\mu$m, and a system FOV of 800$\times$800~$\mu$m$^2$.

To study the polarization effects, a half-wave plate (HWP) was mounted on a motorized rotation stage (K10CR1, Thorlabs) and placed in the illumination beam path right before the illumination objective to vary the polarization direction of linearly polarized light. We first determined the polarization state of the incident laser by adding a polarizer behind the illumination objective and measuring the laser power behind the polarizer. As shown in Fig.~\ref{fig:setup}(a), polarized light with its electric field along the plane of incidence is defined as p-polarization (HWP at $\sim$70$^{\circ}$), while light whose electric field is normal to the plane of incidence is denoted s-polarization (HWP at $\sim$25$^{\circ}$). The geometries of fluorescence emission with respect to these two representative polarization states are sketched in Fig.~\ref{fig:setup}(b). When the orientation of the transition dipole moment (red arrow) is along the electric field of the excitation light, it maximizes the excitation probability and emits the most fluorescence in the plane perpendicular to the dipole moment, thus leading to the anisotropic fluorescence emission. The orthogonal configuration of the excitation and detection axes enhances the presence of anisotropy since it introduces spatial constraints in the collection of the emitted fluorescence.

The laser power used for excitation was measured at the pupil of the illumination objective to be $\sim$0.35~mW for both experiments on the fluorescent microspheres and zebrafish larvae. The power stability of the excitation light was carefully checked before data acquisition. We measured three sequences of the laser powers via rotating the HWP with a step size of 2 degrees. The power data are plotted in Fig.~\ref{fig:setup}(c), where the black line and the red error bar are the mean value and the standard deviation of the laser powers of three measurements, respectively. With the turning of laser polarization, the laser power is very stable with a mean value of 0.346~mW and a standard deviation of 6.29$\times 10^{-4}$~mW. In the polarization-dependent experiment, a fluorescence image was recorded every 2 degrees of HWP rotation with a full range of 90 degrees covering s- and p- polarizations. To collect enough fluorescence to obtain an informative image from a single transversal plane, the exposure time was set as 50~ms. 

The photobleaching effect was examined by continuously irradiating zebrafish larvae with the same laser power of 0.35~mW, which is far from the saturation regime or damage threshold~\cite{Woo:BOE:248408,DiCicco:IOVS:6281-6288}. We recorded the fluorescence images every 5~minutes and show the whole brightness as a function of time in Fig.~\ref{fig:setup}(d). The overall brightness monotonously decreased to 70$\%$ of the initial value after 90~minutes. In our experiments, including the time consumed by the rotation and stabilization of the motorized rotator (maximum velocity: 15$^{\circ}$/s, acceleration speed: 15$^{\circ}$/s), it took less than 1~minute to record 46 images to accomplish one polarization period. Therefore, the fluorescence degradation due to the photobleaching effect within one polarization scan is less than 0.4$\%$ and has little influence on the experimental results.



\section{Results and discussion}

\begin{figure}[htbp]
    \centering
    \includegraphics[width=\columnwidth]{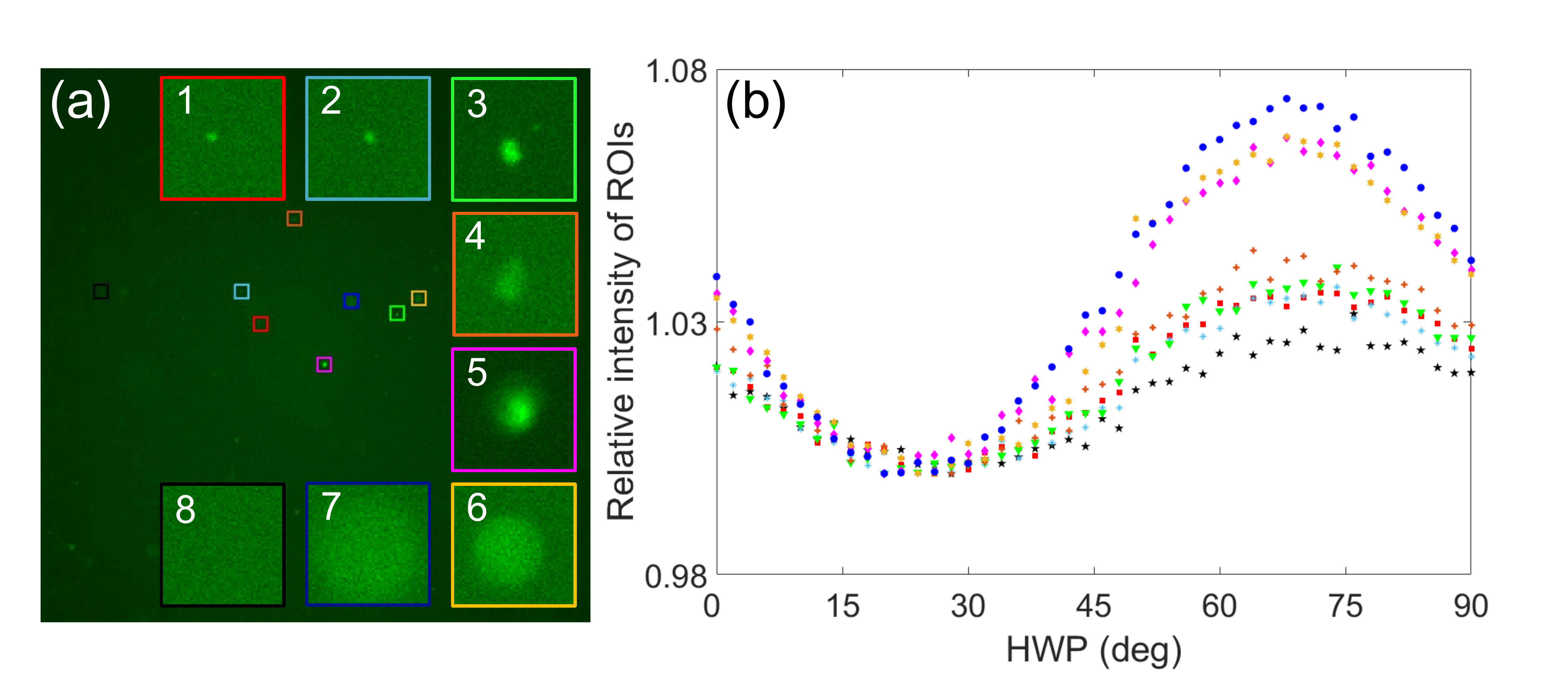}
    \caption{(a) A light-sheet microscopy image recorded from fluorescent microspheres. Insets are the zoom-in images of selected ROIs, where inset 1 and 2 show a single microsphere, inset 3-7 show microsphere clusters with increasing cluster size, and inset 8 shows only the background. (b) The scatter plots of relative intensities as a function of HWP turning angles for all the ROIs. The color of the scatter plot corresponds to the fluorescent microsphere data in the inset of the same outer frame color in the left fluorescent image.}
    \label{fig:polar_microspheres}
\end{figure}

We first tested the effect of laser polarization on the fluorescence emission of microspheres. Fig.~\ref{fig:polar_microspheres}(a) shows a light-sheet fluorescence image of the microspheres. A single microsphere is of diameter 0.5~$\mu$m, occupying 3$\sim$4 pixels. Insets are the zoom-in images of the regions of interest (ROIs), enclosing an area containing either a single microsphere (Inset 1, 2) or a cluster of microspheres (Inset 3-7) or background noise (Inset 8), with sizes of 50 by 50 pixels. The relative intensities of all the ROIs are plotted as a function of HWP tuning angles in Fig.~\ref{fig:polar_microspheres}(b). The color of the scatter plot corresponds to the color of the outer frame of the inset in Fig.~\ref{fig:polar_microspheres}(a). The summation of values of each pixel in the ROI is defined as fluorescence intensity and the relative intensity is calculated as the intensity divided by the minimum intensity in the same plot, which is also referred to as the anisotropy value in this paper. We found that the background noise has the smallest anisotropy value of less than 1.032, which is attributed to the slight difference in the refractive index of agarose under different polarizations. The intensity modulation of the fluorescence emission from microspheres is similar to the one from the background, since the refractive index difference plays the same role in the fluorescence emission from the microspheres by changing the incident light intensity. The anisotropy values of a single microsphere (Inset 1, 2) or cluster of few microspheres (Inset 3, 4, 5) are between 1.032 and 1.045, while for a big cluster of microspheres (Inset 6, 7) are between 1.067 and 1.074. These values indicate the difference in fluorescence intensity due to different polarizations is about 4$\%$ or 7$\%$, meaning that the fluorescence emission hardly depends on the excitation polarization. In our experiment, the microspheres were prepared with agarose, which restricted their movement. The fluorophores embedded in the microspheres are constrained by their bindings and lack the degree of freedom. Therefore, we believe that nearly isotropic fluorescence emission comes from a population of randomly-oriented fluorophores. Despite weak, a single fluorescent microsphere still presents anisotropic fluorescence emission. With more and more carboxylate-modified microspheres bound by chemical bonds, we observed that the anisotropy value increases with the increase of cluster size, showing a moderate preference for light polarization and fluorescence emission direction.

\begin{figure}[htbp]
    \centering
    \includegraphics[width=\columnwidth]{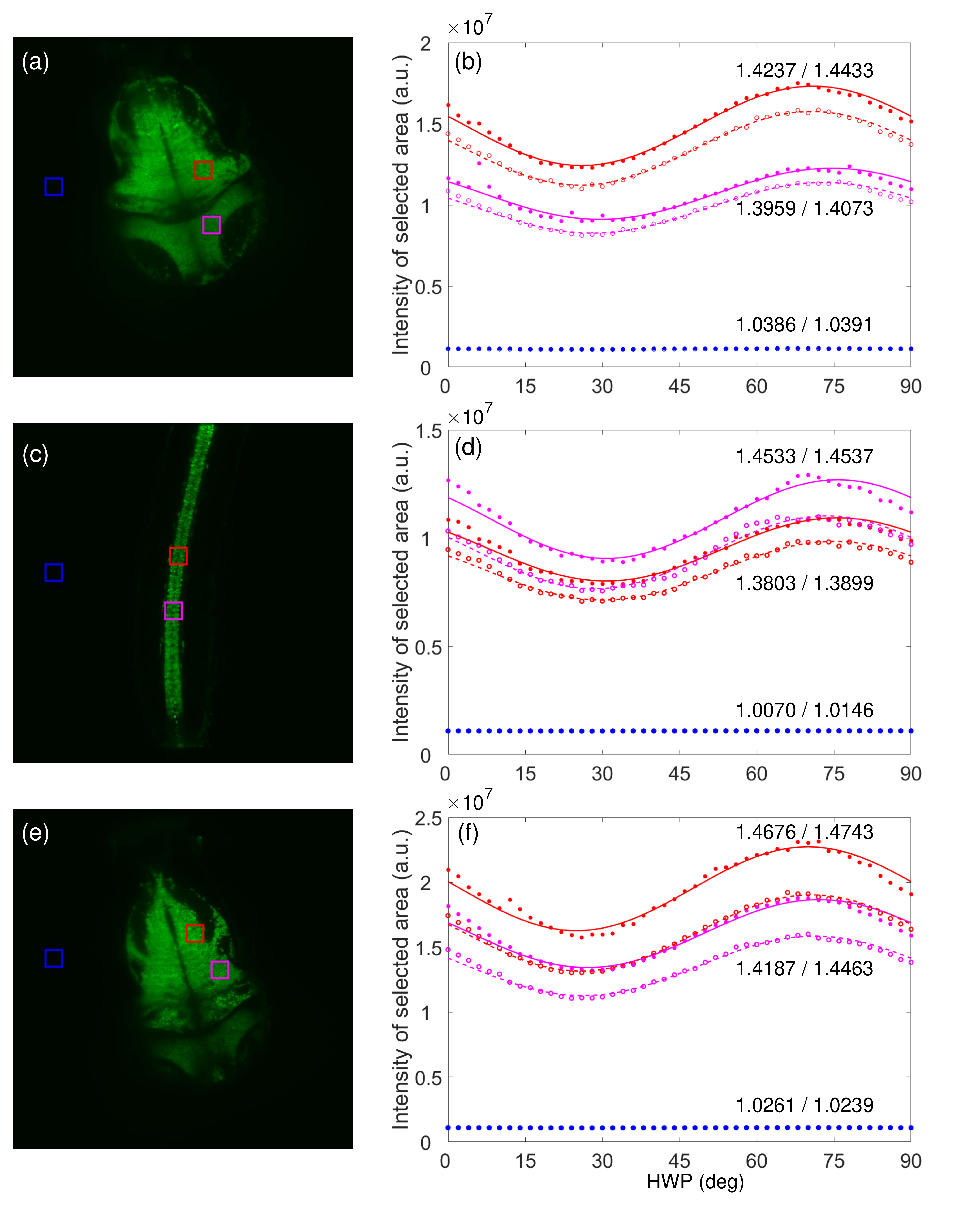}
    \caption{Fluorescence images recorded mainly at (a) midbrain; (c) tail; and (e) hindbrain of zebrafish larvae at p-polarization. (b,~d,~f) The scatter plots of fluorescence intensity as a function of HWP angles for three selected ROIs marked in the left images. The color of the scatter plot corresponds to the color of the square mark. The solid and hollow data points were recorded with the laboratory light on and off, respectively. Among them, the blue data are the background, whose solid and hollow ones are overlapped. The red and magenta data are taken from the emission regions, whose solid and hollow ones are fitted with a cosine square function, plotted in solid and dash lines, respectively. The polarization dependence curves show around 38-47$\%$ higher fluorescence emission at p-polarization for all the cases except the background. 
    }
    \label{fig:polar_zebrafish}
\end{figure}    


We then examined the polarization effect on GCaMP6f-expressed zebrafish larvae utilizing the same self-built LSFM system. GCaMP6f is a genetically encoded fluorescent Ca[2+] indicator, consisting of a circularly permuted green fluorescent protein (GFP) and calmodulin, and expresses under a pan-neuronal promoter. Figure.~\ref{fig:polar_zebrafish}(a),~(c),~and~(e) display the fluorescence images of three different parts of zebrafish larvae. They were all recorded under p-polarization with maximized fluorescence signal. Similarly, we analyzed three ROIs with sizes of 100 by 100 pixels, two from the emission regions and one from the background. The intensity changes as a function of HWP angles within certain ROIs are plotted in Fig.~\ref{fig:polar_zebrafish}(b),~(d),~and~(f) in the same color as the square markers of ROIs in the fluorescence images to the left. The solid and hollow data points are the control results with the laboratory light turned on and off, respectively, during the record of fluorescence images. The anisotropic values are listed vertically in the same order as solid data, the values after the slash are for the hollow data of the same color. As shown in Fig.~\ref{fig:polar_zebrafish}(b,~d,~f), the solid and hollow data points in blue color are overlapped, indicating that the intensity of background noise is unrelated to the laboratory lighting. The intensity gap between the light on and off from the other two ROIs (in color red and magenta) is mainly due to the reflection of the laboratory light from the zebrafish larva surface. Due to both the photoselection effect and the non-coaxial configuration of LSFM, the intensity of fluorescence emission taken at p-polarization is about 38-47$\%$ stronger than the one taken at s-polarization. This phenomenon has also been observed recently by another group~\cite{Vito:BOE:4651-4665}. The sharp sparks, {\it{e.g.}} the magenta solid data point at HWP=6$^{\circ}$ in Fig.~\ref{fig:polar_zebrafish}(b), signify the neuronal signal transmission that happened inside the magenta square during the 50~ms exposure time. These sparks affect the smoothness of the polarization correlation curve but without changing the trend of modulation. We have the same experiments repeated on two other zebrafish larvae and obtain comparable anisotropic emission properties. Considering the particularity of the biological samples, the anisotropy value can vary from 1.3 to 1.5. 

According to the theoretical model, for single-photon excitation, the probability of excitation is proportional to cos$^2\theta$, where $\theta$ is the angle between the transition dipole moment of the molecule and the polarization orientation of the excitation light~\cite{Vinegoni:NP:1472-1497}. We fit the experiment data from the emission regions with a cosine square function of HWP angles in solid and dash lines for light on and off, respectively, as shown in Fig.~\ref{fig:polar_zebrafish}(b,~d,~f). In general, the data points are in good agreement with the theoretical model with the maximum fluorescence intensity $I_\text{p}$ observed at p-polarization and minimum intensity $I_\text{s}$ at s-polarization. However, the amount of fluorescence at s-polarization is more than what we expected. As sketched in Fig.~\ref{fig:setup}(b), when the excitation light is s-polarized, the favored transition dipole moment oriented along the detection axis fluoresces predominantly in the xy-plane. The detection objective used in the experiment has a collecting angle of 37$^{\circ}$ and can only gather a small portion of the fluorescence signal. We infer that the considerable amount of fluorescence at s-polarization is caused by the rotational diffusion of the fluorescent molecules, which is the dominant depolarization process that changes the direction of transition dipole moments. This intensity part is expressed as $I_\text{depol}$, also exists in p-polarization.

The ratio between p- and s-polarization reflects the fraction of the molecules attached by the fluorescence tag with a preferential direction. According to the electric dipole radiation, the intensity observed at angle $\alpha$ is proportional to $sin^2\alpha/r$, where $\alpha$ is the angle between the dipole orientation and the observation point, and $r$ is the distance from the dipole center to the observation point. In our experiment, the intensity signal is then the integral of $sin^2\alpha/r$ with the integration limits corresponding to the 37$^{\circ}$ collection angle. Therefore, for the same amount of fluorophores, the intensity ratio of the two representative schemes in Fig.~\ref{fig:setup}(b) is calculated to be $I_\text{p}/I_\text{s}\sim$2.25. If we consider a simple case that $(I_\text{p}+I_\text{depol})/(I_\text{s}+I_\text{depol})\approx1.4$, then the number of molecules with dipole orientations along with p-polarization $N_\text{p}$, s-polarization $N_\text{s}$ and random orientations $N_\text{depol}$ would satisfy the relation as $2.25N_\text{p}-1.4N_\text{s}=0.4N_\text{depol}$. However, the actual situation of living animals is much more complicated. The fluorescence intensity detected under each linear-polarized light is the superposition of the fluorescence of molecules with different dipole moment orientations. In addition, the resonance effect would become an important consideration when the length of the dipole is comparable to the excitation wavelength.

\begin{figure}[htbp]
    \centering
    \includegraphics[width=\columnwidth]{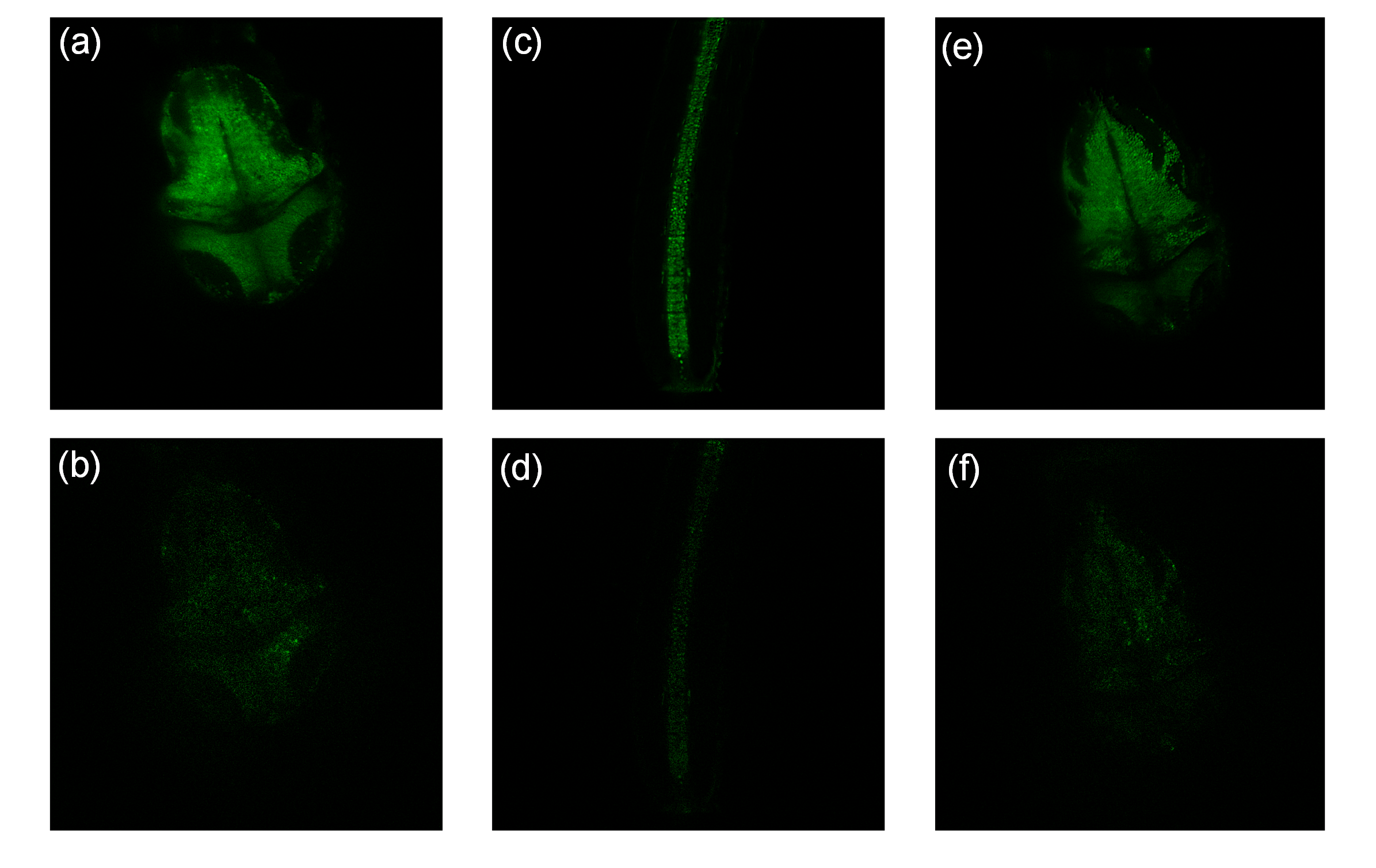}
    \caption{The subtraction of reference fluorescence images from the ones at p-polarization at (a) midbrain; (c) tail; and (e) hindbrain of the zebrafish larvae. The subtraction of reference fluorescence images from the ones at s-polarization at (b) midbrain; (d) tail; and (f) hindbrain.}
    \label{fig:subtraction}
\end{figure}

As we have observed the difference in intensity due to the anisotropic fluorescence, which interests us next is whether this anisotropic fluorescence also exhibits a preference for certain regions within the biological sample. To explore this question, one needs to separate the fluorescence emission according to the excitation polarization. For this purpose, the fluorescence images taken at both s-polarization and p-polarization from the same recording sequence were registered at single-cell resolution to a reference image, which is an image also taken at s-polarization but at a different time. Then, the reference image was subtracted from the registered fluorescence images at two representative polarizations without normalizing the initial intensity. The resulting images for three parts of the larval zebrafish are shown in Fig.~\ref{fig:subtraction}. Since the zebrafish experiments were conducted in vivo, subtraction of s-polarization images and the reference image was performed to provide a comparison to identify that the difference was induced by polarization rather than any occasional biological activity.

The upper and lower rows in Fig.~\ref{fig:subtraction} show the reference image subtracted from p-polarization and s-polarization, respectively. Compared to the original images recorded under p-polarization (see Fig.~\ref{fig:polar_zebrafish}(a,~c,~e)), the subtracted images become flocculent. We note that the enhanced fluorescence due to anisotropic emission can be observed throughout the whole emission region and does not show a preference for specific biological sites. For Fig.~\ref{fig:subtraction}(b,~d,~f), the subtractions of s-polarization images and the reference image, the difference is considered as noise signals, which are uniformly scattered over the whole fluorescence emission region, except for the hot spots/sparks arising from the neuronal transmission.

\begin{figure}[htbp]
    \centering
    \includegraphics[width=\columnwidth]{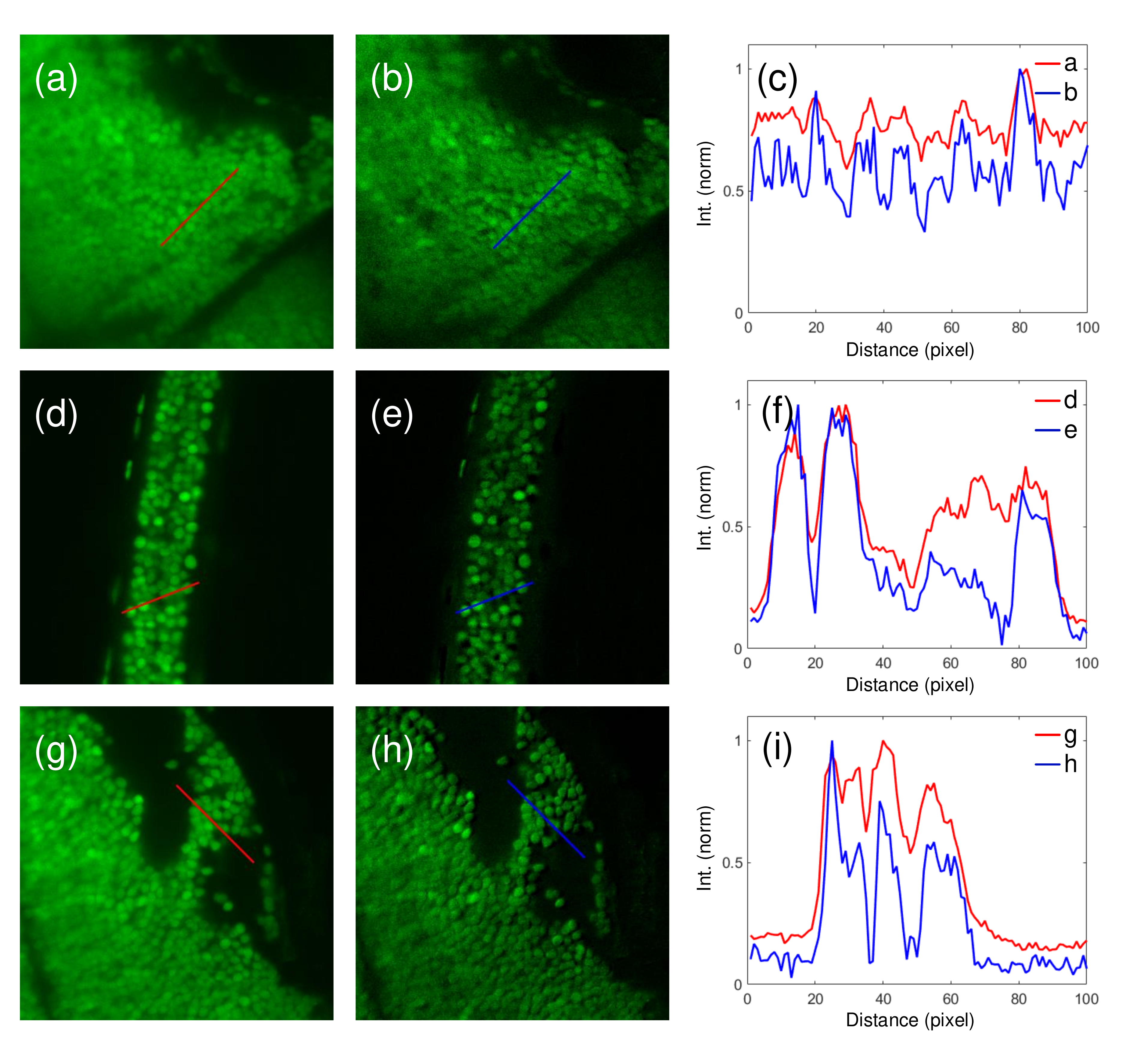}
    \caption{(a,~d,~g) Zoom in images of arbitrary selected areas from Fig.~\ref{fig:polar_zebrafish}(a,~c,~e), size: 400 by 400 pixels; (b,~e,~h) Zoom in images of arbitrary selected areas from Fig.~\ref{fig:subtraction}(a,~c,~e), size: 400 by 400 pixels; (c),~(f) and (i) are the normalized intensity profiles along the blue and red lines in Fig.~\ref{fig:comparison}(a,~b),~(d,~e) and (g,~h), respectively.}
    \label{fig:comparison}
\end{figure}

In order to further dissect the cause of anisotropic fluorescence, arbitrarily selected areas of 400 by 400 pixel size from fluorescence images taken under p-polarizations before and after subtraction are zoomed in Fig. 5, where Fig.~\ref{fig:comparison}(a,~d,~g) are from Fig.~\ref{fig:polar_zebrafish}(a,~c,~e) and Fig.~\ref{fig:comparison}(b,~e,~h) are from Fig.~\ref{fig:subtraction}(a,~c,~e). It can be seen that the biological structures become sharper in the subtracted images. The normalized intensity profiles along the blue and red lines in Fig.~\ref{fig:comparison}(a,~b), (d,~e), and (g,~h) are plotted in Fig.~\ref{fig:comparison}(c),~(f),~and~(i) in the same color, respectively. 
With the normalized peak height, in all three graphs, the fluorescence intensity after subtraction (blue lines) has a much low value at the valley than the fluorescence intensity before subtraction (red lines). Reflecting on the images, this suggests that the anisotropic fluorescence emission due to the polarization on the GCaMP6f-expressed zebrafish larvae mainly occurs within the nerve cells, rather than intercellular or extracellular substances. This observation is rational since the GCaMP is commonly used to measure increases in intracellular Ca$^{2+}$ in neurons as a proxy for neuronal activity. While the fluorescence signal observed in the extracellular space may come from the scattered-fluorescence, auto-fluorescence, Poisson noise, etc, which are usually regarded as the noise signal and degrade the image quality~\cite{Guan:LSA:1-13}. Therefore, based on this phenomenon, a fluorescence image with better contrast and resolution can be obtained by subtracting images under different polarizations. Herein, we use the Michelson formula to quantify the contrast~\cite{Baumgart:OE:21805-21814}. The contrast $C$ is calculated as $C=(I_\text{max}-I_\text{min})/(I_\text{max}+I_\text{min})$, where $I_\text{max}$ and $I_\text{min}$ are the intensity values of each peak and its corresponding valley, respectively. By averaging $C$ values of main peaks in the graph, we obtain the gross contrast with and without differential polarization subtracting method to be 0.30 and 0.13 (Fig.~\ref{fig:comparison}(c)), 0.61 and 0.35 (Fig.~\ref{fig:comparison}(f)), and 0.77 and 0.37 (Fig.~\ref{fig:comparison}(i)), respectively. Hence, an improvement of $\sim$2 times the image contrast is achieved. And this improvement is more significant in biological structure with many cells. Last but not least, in the first 20-pixel distance in Fig.~\ref{fig:comparison}(c), extra biological structures can be resolved into distinct objects. Elimination of depolarized fluorescence emission enables one to distinguish the boundaries of nerve cells more clearly, which is promising for future research with a superior requirement for structural information. 

\section{Conclusion}

We have successfully examined the effects of excitation light polarization on fluorescence emission using self-built light-sheet microscopy. Zebrafish larvae (GCaMP6f-expressed) embedded in low-melting-point agarose have been used as the biological organism to explore the anisotropic fluorescence emission. As expected, the levels of the fluorescence signal follow the cosine square function with respect to the polarization orientation of the excitation light. An approximate 40$\%$ increased fluorescence emission is observed when the excitation light is p-polarized, which principally excites the molecules with transition dipole moments lined up in the direction perpendicular to the detection axis, therefore emitting the most fluorescence toward the direction of the detection objective. This provides a straightforward way to adjust the fluorescence intensity and obtain the maximum signal during the light-sheet imaging experiments. From the ratio of the fluorescence emission of p- and s-polarization, we discuss the orientation of the molecules the fluorescence tag attached to, which would be interesting for biologists to further investigate nerve cells based on the intrinsic properties of the biological structure. We have further analyzed the origin of anisotropic fluorescence emission via image processing. After the registration and subtraction of the fluorescence images of various polarization, we find the anisotropy is driven by the fluorescence signal arising in the nerve cells rather than the extracellular substance. The fluorescence emission from the extracellular substance is comparable under p- and s-polarization, and thus can to a large extent be eliminated through subtraction, leaving a fluorescence image with improved contrast and increased resolution of detached cell structure. We have demonstrated that using this differential polarization subtracting method, one can obtain a clearer structural fluorescence image of twice the contrast and resolve more details, which is propitious for future research relying on microscopy imaging techniques.


\begin{backmatter}
\bmsection{Funding}
This work is supported by the "Strategic Priority Research Program" of the Chinese Academy of Sciences (grant no. XDA16021304) and the "Doctor of entrepreneurship and innovation program" in Jiangsu Province (grant no. JSSCBS20211440).

\bmsection{Acknowledgments}
The authors thank Shan Zhao, affiliated with the Institute of Neuroscience, Chinese Academy of Sciences, for preparing the larval zebrafish samples.  

\bmsection{Disclosures}
The authors declare no conflicts of interest.

\end{backmatter}

\bibliography{LSFM}
\end{document}